\documentclass[aps,twocolumn,prd,epsf]{revtex4}

\newcommand{\be}{\begin{equation}}
\newcommand{\ee}{\end{equation}}
\def\bq{\begin{eqnarray}}
\def\eq{\end{eqnarray}}
\def\beq{\begin{eqnarray}}
\def\eeq{\end{eqnarray}}

\def\ba{\begin{eqnarray}}
\def\ea{\end{eqnarray}}

\usepackage{psfig}
\usepackage{graphicx}
\usepackage{dcolumn}
\usepackage{bm}

\newcommand{\Ph}{\ensuremath{\Phi}}
\newcommand{\Pdot}{\ensuremath{\dot{\Phi}}}

\begin{document}

\title{Naked Singularity formation in scalar field collapse}      

\author{Rituparno Goswami and Pankaj S Joshi}

\address{Tata Institute of Fundamental Research,\\ Homi Bhabha Road,\\ 
Mumbai 400 005, India}

 
\begin{abstract} We construct here a class of collapsing scalar 
field models with a non-zero potential, which result in a naked 
singularity as collapse end state. The weak energy condition is 
satisfied by the collapsing configuration. It is shown that physically 
it is the rate of collapse that governs either the black hole or naked 
singularity formation as the final state for the dynamical evolution. 
It is seen that the cosmic censorship is violated in  
dynamical scalar field collapse.

\end{abstract}
 
 
\maketitle

There has been considerable interest and discussion recently in 
the string theory context in the phenomena of possible naked singularity 
formation, and validity or otherwise of the cosmic censorship conjecture
\cite{horo}. 
These considerations are mainly centered around models 
of four dimensional gravity coupled to a scalar field with potential 
$V(\phi)$, and various modifications and numerical studies of 
such a scenario. Such models satisfy the positive energy theorem, however, 
in general may violate the energy conditions, within an asymptotically 
anti de Sitter framework. While these considerations have provided 
us with some insights, the basic question of cosmic censorship 
in scalar field collapse still remains very much open to further 
analysis.

The issue of naked singularity formation in gravitational
collapse is, in fact, of great interest in gravity physics 
and has been investigated already in considerable detail (see e.g. 
\cite{rev}
and references therein) within the framework of Einstein's gravity.
This is because the occurrence of naked singularities 
will offer us the possibility to observe the quantum gravity 
effects taking place in such visible ultra-strong gravity regions. 
The generic conclusion here is, depending on the nature of the
regular initial data (in terms of the initial distribution of 
matter fields and other collapse parameters) from which the collapse 
evolves, the final outcome is either a black hole or a naked singularity. 
Such studies have involved various forms of matter such as dust, perfect 
fluids, radiation collapse, and general (type I) matter 
fields.

Special importance however, is attached at times to the investigation 
of collapse for a scalar field. This is because one would like to
know if cosmic censorship is necessarily preserved or violated in 
gravitational collapse for fundamental matter fields, which are derived 
from a suitable Lagrangian.

Our purpose here is to construct a class of continual 
collapse models of scalar field with potential, such that no trapped 
surfaces form in the spacetime as the collapse evolves in time, and the 
singularity that develops as collapse endstate is necessarily naked. 
We require that the {\it weak energy condition} 
is preserved through out the collapse, though the pressures would 
be negative closer to the singularity. The 
interior collapsing sphere is matched with a generalized Vaidya exterior 
spacetime to complete the model. We thus see that naked singularities 
are created in scalar field collapse from
generic initial conditions, thus violating the cosmic censorship 
hypothesis.

In order to present a transparent consideration, let us examine a 
spherically symmetric homogeneous scalar field, $\Ph=\Ph(t)$, with a 
potential $V(\Ph)$. This ensures that the interior spacetime must have 
a Friedmann-Robertson-Walker (FRW) metric. Further, let us  
choose the marginally bound $(k=0)$ case. Then the interior metric is of 
the form,
\begin{equation}
ds^2=-dt^2+a^2(t)\left[dr^2+r^2d\Omega^2\right]
\label{eq:FRW}
\end{equation}
where $d\Omega^2$ is the line element on a two-sphere. In this 
co-moving frame, the energy-momentum tensor of the scalar
field is given as,
\begin{equation}
T^t_t=-\rho(t)=-\left[\frac{1}{2}\Pdot^2+V(\Ph)\right]
\label{eq:em1}
\end{equation}
and
\begin{equation}
T^r_r=T_{\theta}^{\theta}=T_\phi^\phi=p(t)=\left[\frac{1}{2}\Pdot^2-V(\Ph)
\right]
\label{eq:em2}
\end{equation}
with all other off-diagonal terms being zero.

It may be noted that the comoving coordinate system we have 
chosen has a particular physical significance as compared to an 
arbitrary system, and the quantities $\rho$ and $p$ are interpreted as
the density and pressure 
respectively of the scalar field. It is then easily seen that the scalar 
field behaves like a perfect fluid, as the radial and tangential
pressures are equal. We take the scalar field to 
satisfy the weak energy condition, that is, the energy density as
measured by 
any local observer be non-negative, and for any timelike vector 
$V^i$, we have,
\begin{equation}
T_{ik}V^iV^k\ge0
\end{equation}
This amounts to,
\begin{equation}
\rho\ge0;\;\; \rho+p\ge0;
\label{eq:wec}
\end{equation}

The dynamic evolution of the system is now determined by 
the Einstein equations, which for the metric (\ref{eq:FRW}) 
become (in the units $8\pi G=c=1$),
\begin{eqnarray}
\rho=\frac{F'}{R^2R'}; && p=\frac{-\dot{F}}{R^2\dot{R}}
\label{eq:ein1}
\end{eqnarray}
\begin{equation}
\dot{R}^2=\frac{F}{R}
\label{eq:ein4}
\end{equation}
Here $F=F(t,r)$ is an arbitrary function, and in spherically symmetric 
spacetimes it has the interpretation of the mass function for the 
cloud, with $F\ge0$. The quantity $R(t,r)=ra(t)$ is the area radius for 
the shell labeled by the comoving coordinate $r$.
In order to preserve the regularity of the initial data,
we have  
$F(t_i,0)=0$, that is, the mass function should vanish at the center 
of the cloud. From equation (\ref{eq:ein1}) it is evident that on any regular
epoch $F\approx r^3$ near the center.

The {\it Klein-Gordon} equation for the scalar field is
given by,
\begin{equation}
\frac{d}{dt}\left[a^3\Ph\right]=-a^3V(\Ph)_{,\Ph}
\label{eq:kg}
\end{equation}
Since our aim here is to construct a continual collapse model,
we consider the class with $\dot{a}<0$, which is the collapse condition 
implying that the area radius of a shell at a constant value of 
comoving radius $r$ decreases monotonically.  
As such there may be classes of solutions where a scalar 
field may 
disperse away also (see e.g.
\cite{chop}).
Our objective, however, is to 
examine here whether the singularities forming in scalar field 
collapse could be naked, or necessarily covered within a black hole, 
and if so under what conditions.
The singularity resulting from continual collapse is 
given by $a=0$, that is when the scale factor vanishes and the area 
radius for all the collapsing shells becomes zero. At the singularity 
we must have $\rho\rightarrow\infty$.

The key factor that decides the visibility or otherwise 
of the singularity is the geometry of trapped surfaces which may
form as the collapse evolves. These are
two-surfaces in the spacetime from which both outgoing and ingoing
wavefronts necessarily converge
\cite{HE}.
The boundary of the
trapped region in a spherically symmetric spacetime 
is given by the equation,
\begin{equation}
F=R
\label{eq:apphor}
\end{equation} 
The spacetime region where the mass function $F$ satisfies $F<R$ is 
not trapped, while $F>R$ describes a trapped region (see e.g.
\cite{JG}).

In terms of the scale variable $a$, the mass function can be written as,
\begin{equation}
F=r^3\left[\frac{1}{3}a^3\left\{\frac{1}{2}\Ph(a)_{,a}^2\dot{a}^2+
V(\Ph(a))\right\}\right]
\label{eq:f2}
\end{equation}
This is because from equation (\ref{eq:ein1}) we can solve for 
the mass function as,
\begin{equation}
F=\frac{1}{3}\rho(t)R^3
\label{eq:f1}    
\end{equation}
From equation (\ref{eq:f1}), we then see that,
\begin{equation}
\frac{F}{R}=\frac{1}{3}\rho(t)r^2a^2
\label{eq:f/r}
\end{equation}
The above relation decides the trapping or otherwise 
of the spacetime as the collapse develops.
We would like to construct and investigate the classes of 
collapse solutions for a scalar field with potential, where the 
trapping is avoided till the singularity formation, thereby
allowing the singularity to be visible. 
Towards such a purpose, consider the class of models 
where {\it near the singularity} the divergence of the density
is given by,
\begin{equation}
\rho(t)\approx\frac{1}{a(t)}
\label{eq:rho1}
\end{equation}
Then using equation 
(\ref{eq:f2}),
we see that the above condition (\ref{eq:rho1}) implies, 
near the singularity,
\begin{equation}
\frac{1}{2}\Ph(a)_{,a}^2\dot{a}^2+V(\Ph(a))=\frac{1}{a}
\label{eq:rho2}
\end{equation}
Now solving the equation of motion (\ref{eq:ein4}) we get, 
\begin{equation}
\dot{a}=-\frac{\sqrt{a}}{\sqrt{3}}
\label{eq:adot}
\end{equation}
The negative sign above implies a collapse scenario where 
$\dot{a}<0$. Using
equations (\ref{eq:adot}) and(\ref{eq:rho2}) in later part of equation 
(\ref{eq:ein1}) we can now solve for $\Ph$ which is given as, 
\begin{equation}
\Ph(a)=-\ln a
\label{eq:phi}
\end{equation}  
We note that as we approach the singularity, $a\rightarrow0$
implies $\Ph\rightarrow\infty$, that is, the scalar field blows up 
at the singularity.
Finally, using equations (\ref{eq:adot}) and (\ref{eq:phi}) in 
equation (\ref{eq:kg}) we can solve for the potential $V$ as,
\begin{equation}
V(\Phi)=\frac{5}{6}e^\Phi
\label{eq:v}
\end{equation}
Thus we see that near the singularity,
\begin{equation}
\rho(t)\approx\frac{1}{a(t)}\; ;\;p(t)\approx-\frac{2}{3a(t)} 
\label{eq:rhop}
\end{equation}
It is seen that in the limit of approach to the singular epoch $(t=t_s)$ 
we get $F/R=0$ for all shells
and there is no trapped surface in the spacetime.

We see in the model above that the weak energy condition is satisfied 
as $\rho>0$ and $\rho+p>0$, though the pressure would be negative. Also, 
it should be noted that the pressure 
does not have to be negative from the initial epoch, because we have 
required the specific behavior of $\rho$ in equation (\ref{eq:rho1}), 
only near the singularity. We can always choose a
$V(\Phi)$ such that at the initial epoch $\frac{1}{2}\Pdot^2>V(\Ph)$,
and then pressure would be positive. But near the singularity $V(\Phi)$
should behave according to equation (\ref{eq:v}) and hence pressure should 
decrease monotonically from the initial epoch and tend to $-\infty$
at the singularity.

If from an epoch $t=t^*$ (or
equivalently for some $a=a^*$) the density starts growing as $a^{-1}$, 
then integrating equation (\ref{eq:adot}) we get the singular epoch as,
\begin{equation}
t_s=t^*+2\sqrt{3}a^*
\label{eq:ts}
\end{equation} 
Thus the collapse reaches the singularity in a finite comoving time,
where the matter energy density as well as the 
Kretschman scalar $\kappa=R^{ijkl}R_{ijkl}$ diverges.
We note that 
from equation of motion (\ref{eq:ein4}) it follows that the metric 
function $a$ is given by,
\begin{equation}
a(t)=\left[\sqrt{a^*}-\frac{1}{2\sqrt{3}}(t-t^*)\right]^2
\label{eq:a1}
\end{equation} 
This completes the interior solution within the collapsing cloud,
providing us with the required construction.

We can see from the above considerations that the 
absence or otherwise of trapped surfaces, and the behaviour of pressure, 
crucially depend on the rate 
of divergence of the density $\rho$ near the singularity. 
To examine this more carefully, let us write near the singularity,
\begin{equation}
\rho=a^{-n};\;\;\;n>0
\label{eq:rhon}
\end{equation}
as we know $\rho(t)$ must diverge as $a(t)$ goes
to zero in the limit of approach to the singularity. 
In that case, solving the Einstein equations gives,
\begin{equation}
p=\frac{n-3}{3}a^{-n}
\label{eq:p2}
\end{equation} 
The corresponding values of $\Ph$ and $V(\Ph)$ are,
\begin{eqnarray}
\Ph=-\sqrt{n}\ln(a);&V(\Ph)=\left(1-\frac{n}{6}\right)e^{\sqrt{n}\Ph}
\label{eq:phi2}
\end{eqnarray} 
Again calculating $F/R$ in this general case, we have,
\begin{equation}
\frac{F}{R}=\frac{1}{3}a^{2-n}
\label{eq:f/r2}
\end{equation} 
Thus we see that for low enough divergences $(0<n<2)$ we have 
no trapped surfaces forming and there are negative pressures near the 
singularity.
For $2\le n<3$, trapped surfaces do form, however pressure still 
remains negative at the singularity. For $n\ge3$ we have $p\ge0$
and there are trapped surfaces in the spacetime as collapse
advances. Conversely,
we can say that non-negative pressure always ensures trapped
surfaces in homogeneous scalar field collapse. Thus, the role of 
the potential $V(\Ph)$ chosen is to control the divergence of 
density near the singularity, which in turn governs the 
development or otherwise of trapped surfaces. Fig.1 shows 
the behaviour of the 
functions $V(\Ph)$ with respect to $\Ph$, for different values of $n$.
It is seen that the naked singularity arises from a non-zero
measure open set of initial conditions $(n<2)$, whereas rest 
of the initial data set produces black hole as final
collapse end state.
\begin{figure}[tb!]
\hspace*{0.5cm}
\psfig {file=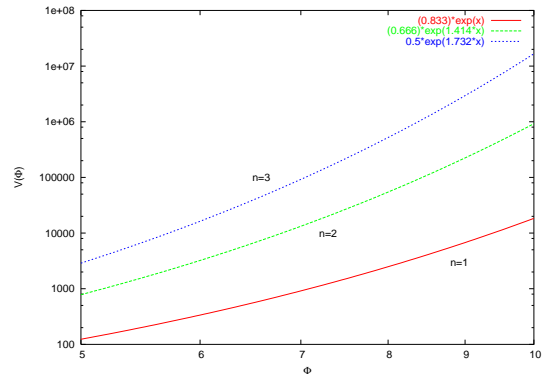,width=7cm,angle=-90}
\caption{$V(\Ph)$ vs $\Ph$ for different $n$. The line $n=2$
separates the black hole and naked singularity regions.}
\end{figure}

To complete the model, we now need to match this
interior to a suitable exterior spacetime. In the following, 
we match this spherical ball of collapsing scalar field
to a generalized Vaidya exterior geometry
\cite{wang}
at the boundary hypersurface $\Sigma$
given by $r=r_b$. Then the metric just inside $\Sigma$ is,
\begin{equation}
ds^2_{-}=-dt^2+a^2(t)\left[dr^2+r_b^2d\Omega^2\right]
\label{eq:metric1}
\end{equation} 
while the metric in the exterior of $\Sigma$ is
\begin{equation}
ds^2_{+}=-\left(1-\frac{2M(r_v,v)}{r_v}\right)dv^2-2dvdr_v+r_v^2d\Omega^2
\label{eq:metric2}
\end{equation} 
where $v$ is the retarded (exploding) null co-ordinate and $r_v$
is the Vaidya radius. Matching the area radius at the boundary we get,
\begin{equation}
r_ba(t)=r_v(v)
\label{eq:radius}
\end{equation} 
Then on the hypersurface $\Sigma$, the interior and exterior metric are
given by,
\begin{equation}
ds^2_{\Sigma-}=-dt^2+a^2(t)r_b^2d\Omega^2
\label{eq:metric3}
\end{equation} 
and
\begin{equation}
ds^2_{\Sigma+}=-\left(1-\frac{2M(r_v,v)}{r_v}+2\frac{dr_v}{dv}\right)dv^2
+r_v^2d\Omega^2
\label{eq:metric4}
\end{equation}
\begin{figure}[tb!]
\hspace*{0.5cm}
\psfig {file=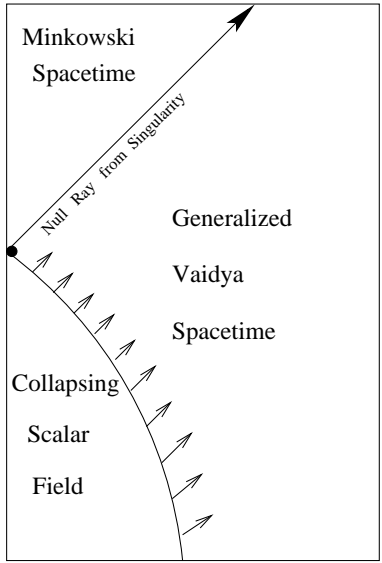,width=5cm}
\caption{A schematic diagram of the complete spacetime.}
\end{figure} 
Matching the first fundamental form on this hypersurface we get,
\begin{eqnarray}
\left(\frac{dv}{dt}\right)_\Sigma=\frac{1}{\sqrt{1-\frac{2M(r_v,v)}{r_v}
+2\frac{dr_v}{dv}}};&\left(r_v\right)_\Sigma=r_ba(t)
\label{eq:match1}
\end{eqnarray} 
To match the second fundamental form (extrinsic curvature)
for interior and exterior metrics, we note that the normal to the
hypersurface $\Sigma$, as calculated from the interior metric, 
is given as,
\begin{equation}
n^i_{-}=\left[0,a(t)^{-1},0,0\right]
\label{eq:n1}
\end{equation} 
and the non-vanishing components of normal derived from the 
generalized Vaidya metric are,
\begin{equation}
n^v_{+}=-\frac{1}{\sqrt{1-\frac{2M(r_v,v)}{r_v}+2\frac{dr_v}{dv}}}
\label{eq:n2}
\end{equation} 
\begin{equation}
n^{r_v}_{+}=\frac{1-\frac{2M(r_v,v)}{r_v}+\frac{dr_v}{dv}}
{\sqrt{1-\frac{2M(r_v,v)}{r_v}+2\frac{dr_v}{dv}}}
\label{eq:n3}
\end{equation} 
Here the extrinsic curvature is defined as,
\begin{equation}
K_{ab}=\frac{1}{2}{\cal L}_{\bf n}g_{ab} 
\label{eq:k1}
\end{equation} 
That is, the second fundamental form is the Lie derivative
of the metric with respect to the normal vector ${\bf n}$.
The above equation is equivalent to,
\begin{equation}
K_{ab}=\frac{1}{2}\left[g_{ab,c}n^c+g_{cb}n^c_{,a}
+g_{ac}n^c_{,b}\right]
\label{eq:k2}
\end{equation} 
Setting $\left[K_{\theta\theta}^{-}-K_{\theta\theta}^{+}\right]_{\Sigma}=0$
on the hypersurface $\Sigma$ we get,
\begin{equation}
r_ba(t)=r_v\frac{1-\frac{2M(r_v,v)}{r_v}+\frac{dr_v}{dv}}
{\sqrt{1-\frac{2M(r_v,v)}{r_v}+2\frac{dr_v}{dv}}}
\label{eq:match2}
\end{equation} 
Simplifying the above equation using equation (\ref{eq:match1}) and
(\ref{eq:ein4}) we get, on the boundary,
\begin{equation}
F(t,r_b)=2M(r_v,v)
\label{eq:match3}
\end{equation} 
Using the above equation and (\ref{eq:match1}) we now get,
\begin{equation}
\left(\frac{dv}{dt}\right)_\Sigma=\frac{1+r_b\dot{a}}{1-\frac{F(t,r_b)}
{r_ba(t)}}
\label{eq:match4}
\end{equation} 
Finally, setting $\left[K_{\tau\tau}^{-}-K_{\tau\tau}^{+}\right]_{\Sigma}=0$,
where $\tau$ is the proper time on $\Sigma$,  
we get,
\begin{equation}
M(r_v,v)_{,r_v}=\frac{F}{2r_ba}+r_b^2a\ddot{a}
\label{eq:match5}
\end{equation} 
Equations (\ref{eq:match3}), (\ref{eq:match4}), (\ref{eq:match5}) along
with (\ref{eq:radius}) completely specify the matching conditions
at the boundary of the collapsing scalar field, as at the boundary
we know the value and the derivatives of the generalized Vaidya mass
function $M(v,r_v)$, which is free otherwise in the
exterior.

Now we can see that at the singular epoch $t=t_s$,
\begin{equation}
\lim_{ r_v\rightarrow 0} \frac{2M(r_v,v)}{r_v}\rightarrow 0
\label{eq:match6}
\end{equation}
Thus the exterior metric along with the singularity smoothly 
transform to,
\begin{equation}
ds=-dv^2-2dvdr_v+r_v^2d\Omega^2
\label{eq:metric5}
\end{equation} 
The above metric describes a Minkowski spacetime in retarded null 
coordinate.
Hence we see that the exterior generalized Vaidya metric, together
with the singular point at $(t_s,0)$, 
can be smoothly extended to the Minkowski spacetime as the collapse 
completes.

From equation (\ref{eq:match4}) it is seen that the 
generalized Vaidya geodesic, which emerges from the singularity 
before it evaporates into free space, is null. It follows that 
non-spacelike trajectories can come out from the singularity 
that develops as the collapse end point. Hence naked singularity
is produced in the collapse of scalar field with potential as 
considered here, for a non-zero measure set of initial conditions, 
and that the occurrence of trapped surfaces in the spacetime is 
avoided. While we have considered here for clarity a homogeneous 
field, the work on inhomogeneous generalization will be 
reported elsewhere.

\end{document}